\newcommand{\CIO}{Cu$_2$IrO$_3$}
\newcommand{\NIO}{Na$_2$IrO$_3$}
\newcommand{\LIO}{Li$_2$IrO$_3$}

\newcommand{\CLSO}{Cu$_{1.5}$Li$_{0.5}$SnO$_3$}
\newcommand{\CNSO}{Cu$_{1.5}$Na$_{0.5}$SnO$_3$}

\newcommand{\C}{$^\circ$C}
\newcommand{\msr}{$\mathrm{\mu}$SR}
\newcommand{\us}{$\mathrm{\mu s}^{-1}$}
\newcommand{\fast}{$\lambda_{\textrm{fast}}$}
\newcommand{\slow}{$\lambda_{\textrm{slow}}$}
\documentclass{nature}

\usepackage[version=3]{mhchem} % Formula subscripts using \ce{}f
\usepackage[T1]{fontenc}       % Use modern font encodings
\usepackage[version=3]{mhchem} % Formula subscripts using \ce{}
\usepackage[T1]{fontenc}       % Use modern font encodings
\usepackage{graphicx}% Include figure files
\usepackage{dcolumn}% Align table columns on decimal point
\usepackage{bm}% bold math
\usepackage{xcolor}% text color
\usepackage{amsmath}% math
\usepackage{hyperref}% add hypertext capabilities

\title{Competition between static and dynamic magnetism in the Kitaev spin liquid material \CIO}

\author{Eric~M.~Kenney$^1$, Carlo~U.~Segre$^2$, William~Lafargue-Dit-Hauret$^3$, Oleg~I.~Lebedev$^4$, Mykola~Abramchuk$^1$, Adam~Berlie$^5$, Stephen~P.~Cottrell$^5$, Gediminas~Simutis$^6$, Faranak~Bahrami$^1$, Natalia~E.~Mordvinova$^4$, Jessica.~L.~McChesney$^7$, Gilberto~Fabbris$^7$, Daniel~Haskel$^7$, Xavier~Rocquefelte$^3$, Michael~J.~Graf$^1$, \& Fazel~Tafti$^1$}

\begin{document}
%"C:\Program Files\Sublime Text 3\sublime_text.exe" -n%l "%f"
\maketitle

\begin{affiliations}
 \item Department of Physics, Boston College, Chestnut Hill, MA 02467, USA
 \item Department of Physics \& CSRRI, Illinois Institute of Technology, Chicago, IL 60616, USA
 \item Univ Rennes, ENSCR, CNRS, ISCR (Institut des Sciences Chimiques de Rennes) - UMR 6226, F-35000 Rennes, France
 \item Laboratoire CRISMAT, ENSICAEN-CNRS UMR6508, 14050 Caen, France
 \item ISIS Neutron and Muon Source, Science and Technology Facilities Council, Rutherford Appleton Laboratory, Didcot, OX11~0QX, United Kingdom
 \item Laboratory for Muon Spin Spectroscopy, Paul Scherrer Institute, 5232 Villigen PSI, Switzerland
 \item Advanced Photon Source, Argonne National Laboratory, Argonne IL 60439, USA

\end{affiliations}

%%%%%%%%%%%%%%%%%%%%%%%%%%%%%%%%%%%%%%%%%%%%%%%%%%%%%%%%%%%%%%%%%%%%%%%%%%%%%%%%%%%%%%%%%%%%%%%%%%%%%%%%%%%%%%%%%%%%%%%%%%%
%%%%%%%%%%%%%%%%%%%%%%%%%%%%%%%%%%%%%%%%%%%%%%%%%%%%%%%%% Abstract %%%%%%%%%%%%%%%%%%%%%%%%%%%%%%%%%%%%%%%%%%%%%%%%%%%%%%%%
%%%%%%%%%%%%%%%%%%%%%%%%%%%%%%%%%%%%%%%%%%%%%%%%%%%%%%%%%%%%%%%%%%%%%%%%%%%%%%%%%%%%%%%%%%%%%%%%%%%%%%%%%%%%%%%%%%%%%%%%%%%
\pagebreak
\begin{abstract}
%For Nature, the abstract is really an introductory paragraph set
%in bold type.  This paragraph must be ``fully referenced'' and
%less than 180 words for Letters.  This is the thing that is
%supposed to be aimed at people from other disciplines and is
%arguably the most important part to getting your paper past the
%editors.  End this paragraph with a sentence like ``Here we
%show...'' or something similar.
%**
%%
Anyonic excitations emerging from a Kitaev spin liquid can form a basis for quantum computers~\cite{kitaev_anyons_2006,kitaev_fault-tolerant_2003}.
Searching for such excitations motivated intense research on the honeycomb iridate materials~\cite{singh_relevance_2012,choi_spin_2012,chaloupka_zigzag_2013,kimchi_kitaev-heisenberg_2014,hwan_chun_direct_2015,winter_challenges_2016,mehlawat_heat_2017,kitagawa_spinorbital-entangled_2018,slagle_theory_2018,simutis_chemical_2018,plumb_rucl3_2014,sears_phase_2017,takayama_hyperhoneycomb_2015,veiga_pressure_2017,modic_realization_2014}.
However, access to a spin liquid ground state has been hindered by magnetic ordering~\cite{chaloupka_zigzag_2013}.
\CIO\ is a new honeycomb iridate without thermodynamic signatures of a long-range order~\cite{abramchuk_cu2iro3:_2017}.
Here, we use muon spin relaxation to uncover the magnetic ground state of \CIO.
We find a two-component depolarization with slow and fast relaxation rates corresponding to distinct regions with dynamic and static magnetism, respectively.
X-ray absorption spectroscopy and first principles calculations identify a mixed copper valence as the origin of this behavior.
Our results suggest that a minority of Cu$^{2+}$ ions nucleate regions of static magnetism whereas the majority of Cu$^+$/Ir$^{4+}$ on the honeycomb lattice give rise to a Kitaev spin liquid.
\end{abstract}

%%%%%%%%%%%%%%%%%%%%%%%%%%%%%%%%%%%%%%%%%%%%%%%%%%%%%%%%%%%%%%%%%%%%%%%%%%%%%%%%%%%%%%%%%%%%%%%%%%%%%%%%%%%%%%%%%%%%%%%%%%%
%%%%%%%%%%%%%%%%%%%%%%%%%%%%%%%%%%%%%%%%%%%%%%%%%%%%%%% Introduction %%%%%%%%%%%%%%%%%%%%%%%%%%%%%%%%%%%%%%%%%%%%%%%%%%%%%%
%%%%%%%%%%%%%%%%%%%%%%%%%%%%%%%%%%%%%%%%%%%%%%%%%%%%%%%%%%%%%%%%%%%%%%%%%%%%%%%%%%%%%%%%%%%%%%%%%%%%%%%%%%%%%%%%%%%%%%%%%%%
% Then the body of the main text appears after the intro paragraph.
% Figure environments can be left in place in the document.
% \verb|\includegraphics| commands are ignored since Nature wants
% the figures sent as separate files and the captions are
% automatically moved to the end of the document (they are printed
% out with the \verb|\end{document}| command. However, tables must
% be manually moved to the end of the document, after the addendum.
\pagebreak
%\section*{Introduction}
%%
Long-range magnetic order is the natural ground state of an interacting electron system.
Magnetic frustration is capable of disrupting the order and establishing a highly entangled ground state with non-local excitations known as a quantum spin liquid~\cite{savary_quantum_2017}.
Among various spin liquid proposals, the Kitaev model has unique appeal because it offers an exact solution to a simple Hamiltonian $\left(H_{ij}=-\sum_{\gamma}K_{\gamma} S_i^\gamma S_j^\gamma\right)$ of spin-$1/2$ particles with bond dependent ferromagnetic coupling $\left(K_\gamma\right)$~\cite{kitaev_anyons_2006}.
The index $\gamma$ corresponds to three inequivalent bonds at $120^\circ$ on a honeycomb lattice.
Two alkali iridates, \LIO\ and \NIO, were the first proposed Kitaev materials based on their honeycomb lattice structures that accommodate Ir$^{4+}$ ions with pseudospin-1/2 ($J_{\textrm{eff}}=1/2$)~\cite{singh_antiferromagnetic_2010,chaloupka_kitaev-heisenberg_2010,singh_relevance_2012,choi_spin_2012,chaloupka_zigzag_2013,majumder_breakdown_2018}.
Despite satisfying the basic assumptions of a Kitaev model, both compounds exhibited antiferromagnetic ordering with sharp peaks in both DC-magnetization and heat capacity at $15$~K~\cite{choi_spin_2012,mehlawat_heat_2017}.
Further investigations on the honeycomb~\cite{plumb_rucl3_2014,sears_phase_2017}, hyperhoneycomb~\cite{takayama_hyperhoneycomb_2015,veiga_pressure_2017}, and harmonic honeycomb~\cite{modic_realization_2014} materials revealed the presence of a Heisenberg interaction ($J$) and a symmetric off-diagonal interaction ($\Gamma$) in the modified Hamiltonian of Kitaev materials~\cite{rau_generic_2014,hu_first-principles_2015}:
\begin{equation}
\mathcal{H}=\sum\limits_{<i,j>,\gamma\neq \alpha,\beta} \left[ -K_{\gamma}S_i^{\gamma}S_j^{\gamma} + J \textbf{S}_i  \cdot \textbf{S}_j+ \Gamma \left( S_i^{\alpha}S_i^{\beta} + S_i^{\beta}S_i^{\alpha}\right) \right]
\end{equation}
The search for a Kitaev material with a negligible Heisenberg interaction and without a long-range order has recently lead to a new honeycomb copper iridate, \CIO~\cite{abramchuk_cu2iro3:_2017}.
Despite having a similar magnetic moment and Curie-Weiss temperature as the alkali iridates, \CIO\ barely revealed a small peak in DC-magnetization at $2$~K and a broad hump in the heat capacity~\cite{abramchuk_cu2iro3:_2017}.
These results indicated short-range correlations and suggested proximity to the Kitaev spin liquid phase.
A spin liquid ground state is expected to exhibit dynamical local fields without long-range ordering.
In this letter, we use muon spin relaxation (\msr) as a direct probe of local magnetic fields and provide compelling evidence for a Kitaev spin liquid phase in \CIO.
Furthermore, our \msr\ results reveal a competition between dynamic and static magnetism in distinct volumes in the ground state.
The source of such behavior is traced to a mixed valence of Cu$^+$/Cu$^{2+}$ by X-ray absorption spectroscopy and first-principles calculations.
%%

%%%%%%%%%%%%%%%%%%%%%%%%%%%%%%%%%%%%%%%%%%%%%%%%%%%%%%%%%%%%%%%%%%%%%%%%%%%%%%%%%%%%%%%%%%%%%%%%%%%%%%%%%%%%%%%%%%%%%%%%%%%
%%%%%%%%%%%%%%%%%%%%%%%%%%%%%%%%%%%%%%%%%%%%%%%%%%%%%%%%%% SECTION %%%%%%%%%%%%%%%%%%%%%%%%%%%%%%%%%%%%%%%%%%%%%%%%%%%%%%%%
%%%%%%%%%%%%%%%%%%%%%%%%%%%%%%%%%%%%%%%%%%%%%%%%%%%%%%%%%%%%%%%%%%%%%%%%%%%%%%%%%%%%%%%%%%%%%%%%%%%%%%%%%%%%%%%%%%%%%%%%%%%

%\pagebreak
%\section*{Muon spin relaxation}
%%%%%%%%%%%%%%%%%%%%%%%%%%%%%%%%%%%%%%%%%%%%%%%%%%%%%%%% FIGURE 1 %%%%%%%%%%%%%%%%%%%%%%%%%%%%%%%%%%%%%%%%%%%%%%%%%%%%%%%%%
\begin{figure}
\includegraphics[width=\textwidth]{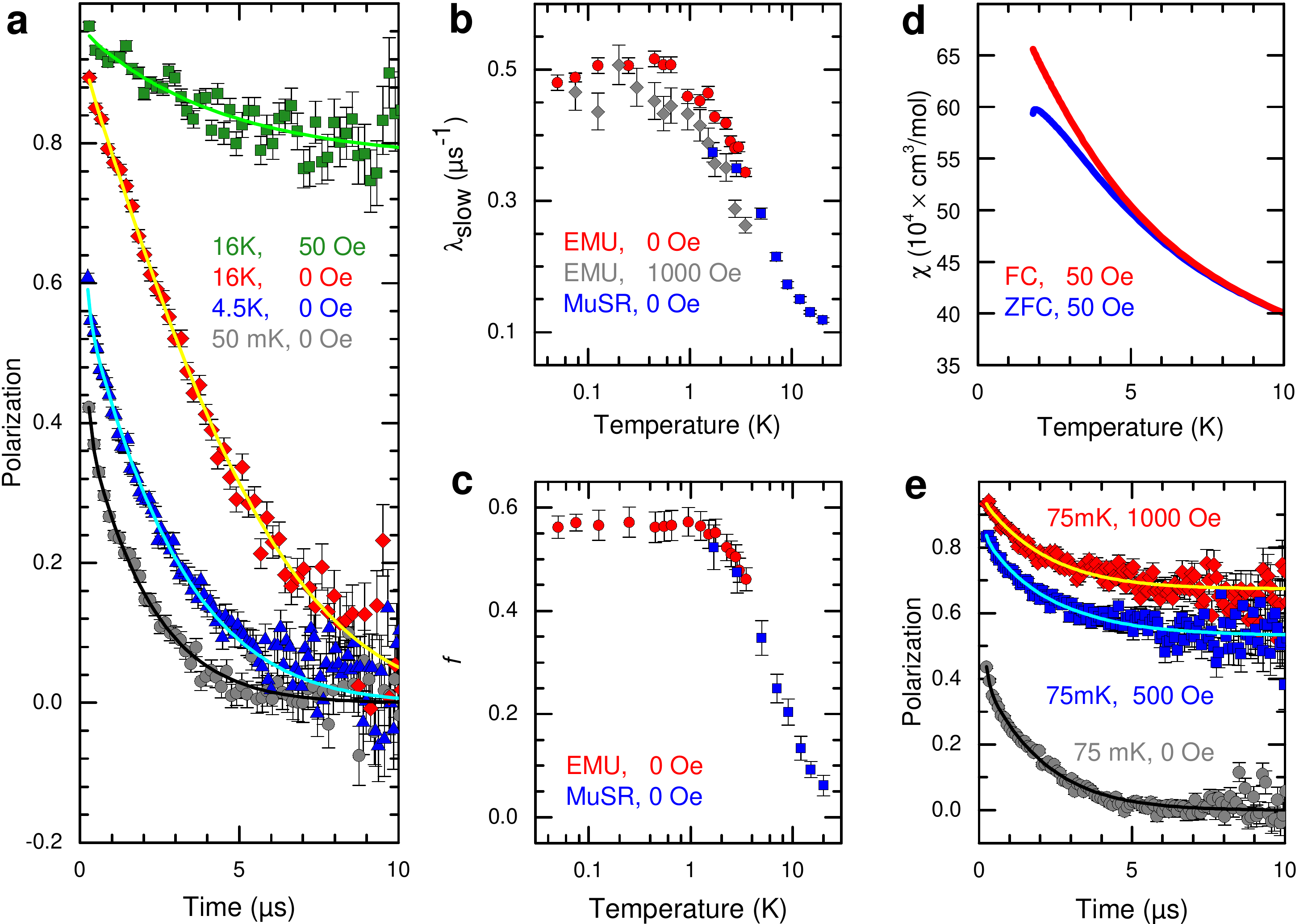}
\caption{\label{MUSR}
\textbf{$\vert$~\msr\ data.}
\small{
\textbf{a,} Representative zero field (ZF) spectra obtained at $16$~K (red diamonds), $4.5$~K (blue triangles), and $0.05$~K (gray circles) as well as longitudinal field (LF) spectrum at $16$~K and $50$~Oe (green squares).
Continuous lines are fits to the data.
Supplementary \msr\ data are presented in Fig.~S1.
%All data in this figure were obtained by the EMU detector.
%
%Similar spectra and fits from the MuSR detector are presented in Supplementary Fig.~S1.
%
\textbf{b,} Temperature dependence of the slow depolarization rate $\lambda_{\textrm{slow}}$ shows a plateau below 2~K at both ZF and LF of $1000$~Oe with data extending over two decades of temperature from $20$ to $0.05$~K.
\textbf{c,} Temperature dependence of the fast depolarization fraction $f$ shows a plateau below $2$~K.
\textbf{d,} DC magnetic susceptibility shows a small peak at $2$~K and a splitting between field-cooled (FC) and zero-field-cooled (ZFC) at $10$~K.
\textbf{e,} \msr\ spectra at $75$~mK in several longitudinal fields show a persistent slow depolarization and a vanishing fast depolarization component.
}
}
\end{figure}
%%%%%%%%%%%%%%%%%%%%%%%%%%%%%%%%%%%%%%%%%%%%%%%%%%%%%%%%%%%%%%%%%%%%%%%%%%%%%%%%%%%%%%%%%%%%%%%%%%%%%%%%%%%%%%%%%%%%%%%%%%%

%%
In \msr, spin polarized positive muons are implanted in the sample, and the time evolution of the muon spin polarization in the local magnetic field is traced upon accumulating several million muon decay events.
In Fig~\ref{MUSR}a, we show three muon polarization spectra in zero applied field (ZF) at $16$, $4.5$, and $0.05$~K, and one spectrum at $16$~K in a $50$~Oe applied field parallel to the initial muon polarization (longitudinal field, or LF).
The ZF spectra at all temperatures are described by
\begin{equation}
\label{musrfit}
P(t)=G_{KT}(t)\left[(1-f)\exp({-\lambda_{\textrm{slow}}t)} + f\exp({-\lambda_{\textrm{fast}}t}) \right]
\end{equation}
where $G_{KT}(t)$ is the Gaussian Kubo-Toyabe function describing depolarization by quasi-static randomly oriented magnetic moments~\cite{Yaouanc_muon_2011} according to $G_{\textrm{KT}}(t)=\frac{1}{3}+\frac{2}{3}(1-\Delta^2t^2)\exp({-\frac{1}{2}\Delta^2t^2})$.
Fits at $16$~K yield $\Delta=0.11$~\us, a typical rate for depolarization by nuclear moments~\cite{kalvius__2014}.
As expected, this relaxation channel is largely suppressed by a weak LF of $50$~Oe (Fig.~\ref{MUSR}a).
The slow and fast exponential decays (\slow\ and \fast) represent a two-component electronic spin contribution to the muon depolarization, and $f$ is the fraction of the signal associated with the fast decay.
We will show below that \slow\ and \fast\ correspond to muons depolarizing in regions of dynamic and static magnetism, respectively.
In Fig.~\ref{MUSR}a, the fast relaxation is primarily observed as a missing polarization at $t<0.2~\mu$s which is outside the bandwidth of the pulsed muon facility.
However, enough of the fast relaxation tail leaks into the spectra in Fig.~\ref{MUSR}a to fit its contribution with a temperature independent relaxation rate $\lambda_{\textrm{fast}}=9(3)$~\us.
A pulsed muon source is particularly suitable to characterize the slow mode with relaxtion rate $\lambda_{\textrm{slow}}=0.48(1)$~\us at $50$~mK (Fig.~\ref{MUSR}b) which is $18$ times slower than \fast.
Temperature dependences of \slow\ and $f$ are shown in Fig.~\ref{MUSR}b,c.
The slow and fast modes grow rapidly below $10$~K.
This onset of magnetism correlates with the temperature at which the field-cooled (FC) and zero-field-cooled (ZFC) susceptibility curves deviate (Fig.~\ref{MUSR}d).
With further decreasing temperature, both \slow\ and $f$ form plateaus below $T=2$~K (Fig.~\ref{MUSR}b,c).
The onset of a plateau in $f$ coincides with a small peak in the ZFC susceptibility (Fig.~\ref{MUSR}d), suggesting the presence of frozen spins in a fraction of the sample volume.
Field dependence of \msr\ can be used to probe the dynamics of the slow and fast modes.
Figure~\ref{MUSR}e shows that the application of a $1000$~Oe LF restores the missing polarization from the fast relaxing muons, indicating the fast relaxation is caused by static local fields that are significantly less than $1000$~Oe.
In contrast, relaxation of the slow component appears to be due to dynamic rather than static local fields.
Because \slow\ $\ll$ \fast, if the local fields were static for slow relaxing muons, we would expect the slow channel to also be suppressed by the $1000$~Oe LF.
Indeed, if the slow relaxation were caused by a static field, the magnitude of that field would be approximated by $B_i=2\pi \lambda_{\textrm{slow}}/\gamma_{\mu}=37$~Oe $\ll$ $1000$~Oe ($\gamma_\mu/2\pi = 135.5$~MHzT$^{-1}$ is the muon gyromagnetic ratio).
The nearly unchanged relaxation rate and amplitude of the slow mode in $1000$~Oe LF (Fig.~\ref{MUSR}b,e) demonstrate that it is caused by fluctuating local fields.
Therefore, we ascribe \fast\ to muons depolarizing in static magnetic domains, and \slow\ to muons depolarizing in distinct regions with spin-liquid-like fluctuating local fields.
The observation of a slight decrease in amplitude of the slow depolarization in Fig.~\ref{MUSR}e, in contrast to the nearly complete suppression of the fast mode suggests that dynamic and static magnetism do not coexist, but rather compete with one another.
The dynamic component is consistent with theoretical predictions of a Kitaev spin liquid in honeycomb iridates~\cite{chaloupka_kitaev-heisenberg_2010,rau_generic_2014,yamaji_first-principles_2014,slagle_theory_2018} but the source of static magnetism is unclear.
Next, we use spectroscopic techniques to clarify this.
%%

%%%%%%%%%%%%%%%%%%%%%%%%%%%%%%%%%%%%%%%%%%%%%%%%%%%%%%%%%%%%%%%%%%%%%%%%%%%%%%%%%%%%%%%%%%%%%%%%%%%%%%%%%%%%%%%%%%%%%%%%%%%
%%%%%%%%%%%%%%%%%%%%%%%%%%%%%%%%%%%%%%%%%%%%%%%%%%%%%%%%%% SECTION %%%%%%%%%%%%%%%%%%%%%%%%%%%%%%%%%%%%%%%%%%%%%%%%%%%%%%%%
%%%%%%%%%%%%%%%%%%%%%%%%%%%%%%%%%%%%%%%%%%%%%%%%%%%%%%%%%%%%%%%%%%%%%%%%%%%%%%%%%%%%%%%%%%%%%%%%%%%%%%%%%%%%%%%%%%%%%%%%%%%
%\pagebreak
%\section*{X-ray absorption spectroscopy}

%%%%%%%%%%%%%%%%%%%%%%%%%%%%%%%%%%%%%%%%%%%%%%%%%%%%%%%% FIGURE 2 %%%%%%%%%%%%%%%%%%%%%%%%%%%%%%%%%%%%%%%%%%%%%%%%%%%%%%%%%
\begin{figure}
\includegraphics[width=\textwidth]{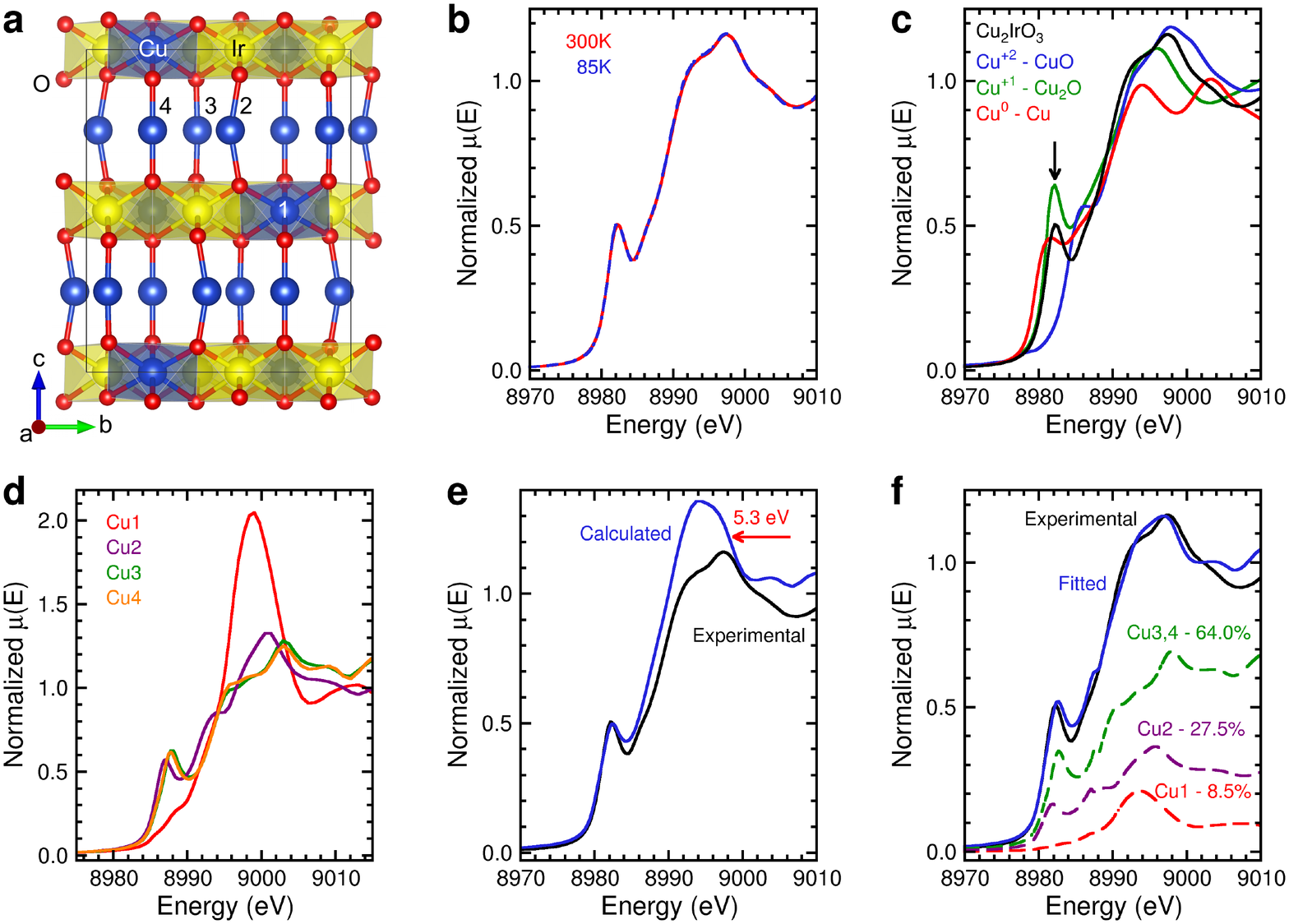}
\caption{\label{XANES}
\textbf{$\vert$~XANES data.}
\small{
\textbf{a,} A Unit cell of \CIO\ viewed down the $a$--axis with four distinct copper sites.
Cu1 in octahedral coordination is within the honeycomb layers whereas Cu2, Cu3, and Cu4 in dumbbell coordination are between the layers.
\textbf{b,} Normalized absorption coefficient plotted as a function of energy in \CIO.
$\mu(E)$ curves are identical at 85 and 300~K.
\textbf{c,} Comparing $\mu(E)$ between \CIO\ and three standard references.
\textbf{d,} Calculated absorption edge of Cu1 to Cu4 using the FEFF software.
\textbf{e,} Absorption spectrum of Cu $K$-edge is calculated by summing over the partial contributions from Cu1 to Cu4 with equal weights.
The calculated signal is shifted by 5.3 eV to match the experimental data with acceptable but not perfect agreement.
\textbf{f,} A fit is made to the experimental XANES data where the weight of each partial contributions is a free parameter.
The resulting weights for Cu1 to Cu4 are reported.
Cu3 and Cu4 have the same weight.
}
}
\end{figure}
%%%%%%%%%%%%%%%%%%%%%%%%%%%%%%%%%%%%%%%%%%%%%%%%%%%%%%%%%%%%%%%%%%%%%%%%%%%%%%%%%%%%%%%%%%%%%%%%%%%%%%%%%%%%%%%%%%%%%%%%%%%

%%
Charge neutrality in \CIO\ dictates conjugate oxidation states of either Cu$^+$ and Ir$^{4+}$, or Cu$^+$/Cu$^{2+}$ and Ir$^{3+}$.
Cu$^+$ $\left[3d^{10}\right]$ is nonmagnetic whereas Cu$^{2+}$ $\left[3d^9\right]$ is magnetic with $S=1/2$.
Ir$^{3+}$ $\left[5d^6\right]$ is nonmagnetic whereas Ir$^{4+}$ $\left[5d^5\right]$ is magnetic with $J_{\textrm{eff}}=1/2$ due to one hole in the $t_{2\textrm{g}}$ level~\cite{kim_novel_2008}.
Each unit cell of \CIO\ (Fig.~\ref{XANES}a) contains three copper sites between the layers (Cu2,3,4) in a dumbbell coordination and one copper site (Cu1) within the honeycomb layers in an octahedral coordination~\cite{abramchuk_cu2iro3:_2017}.
The typical coordination for Cu$^+$ is linear (dumbbells) and for Cu$^{2+}$ is square planar.
An octahedral environment can accommodate both Cu$^+$ and Cu$^{2+}$. %~\cite{shannon_revised_1976}.
Based on this argument we expect at least $75\%$ of Cu$^+$ in \CIO.
X-ray absorption near edge spectroscopy (XANES) is a powerful tool to probe oxidation states.
Our XANES data in Fig~\ref{XANES}b show identical normalized absorption coefficients $\mu(E)$ for Cu $K$-edge at 300 and 85~K confirming a temperature independent ratio Cu$^+$/Cu$^{2+}$ (see Fig.~S2 for Ir $L_3$-edge).
Figure~\ref{XANES}c compares the Cu $K$-edge in \CIO\ at room temperature to Cu, CuO, and Cu$_2$O.
The close similarity with Cu$_2$O indicates a majority of Cu$^+$.
We calculated $\mu(E)$ for the individual sites, Cu1 to Cu4, using the FEFF~8.40 code~\cite{rehr_ab_2009} based on the crystallographic data.
The results in Fig.~\ref{XANES}d show that Cu1 has a spectrum different from Cu2,3,4 as expected from the coordination environments.
Specifically, the edge for Cu1 is shifted to higher energy than the others, indicating a probable Cu$^{2+}$ state.
Since all copper sites in \CIO\ have the same Wyckoff multiplicity~\cite{abramchuk_cu2iro3:_2017}, it is conceivable to reproduce the experimental curve by adding the four partial contributions in Fig.~\ref{XANES}d with equal weight ($25\%$).
The resulting curve in Fig.~\ref{XANES}e shows a mild disagreement with the experimental data.
Specifically, the contribution from Cu1 (nominally Cu$^{2+}$) appears to be overestimated.
The experimental data can be more precisely fit to a weighted sum of partial $\mu(E)$ contributions as reported on Fig.~\ref{XANES}f.
According to this analysis, we estimate $8.5\%$ Cu$^{2+}$ content which means the honeycomb layers contain $1/3$ Cu$^{2+}$ $\left(\frac{8.5\%}{25\%}\right)$ and $2/3$ Cu$^+$.
This is only a rough estimate because we do not know the detailed structure of $\mu(E)$ for Cu$^+$ in octahedral coordination.
Analysis of XANES data from Cu $L_{2,3}$-edges in the Supplementary Fig.~S3 yields an average Cu$^{2+}$ content of $13\%$ which means the honeycomb layers contain $1/2$ Cu$^{2+}$ $\left(\frac{13\%}{25\%}\right)$ and $1/2$ Cu$^+$.
These results are substantiated by self-consistent DFT calculations in the Supplementary Fig.~S4 where the spectroscopic data are best reproduced using $12\%$ Cu$^{2+}$ content.
The spin-1/2 Cu$^{2+}$ ions can nucleate regions of static magnetism within each honeycomb layer giving rise to a fast depolarization of muons (\fast).
Outside these regions, the Cu$^+$/Ir$^{4+}$ combination gives rise to a spin liquid phase with dynamical local fields giving rise to a slow depolarization of muons (\slow).
%%

%%%%%%%%%%%%%%%%%%%%%%%%%%%%%%%%%%%%%%%%%%%%%%%%%%%%%%%%%%%%%%%%%%%%%%%%%%%%%%%%%%%%%%%%%%%%%%%%%%%%%%%%%%%%%%%%%%%%%%%%%%%
%%%%%%%%%%%%%%%%%%%%%%%%%%%%%%%%%%%%%%%%%%%%%%%%%%%%%%%%%% SECTION %%%%%%%%%%%%%%%%%%%%%%%%%%%%%%%%%%%%%%%%%%%%%%%%%%%%%%%%
%%%%%%%%%%%%%%%%%%%%%%%%%%%%%%%%%%%%%%%%%%%%%%%%%%%%%%%%%%%%%%%%%%%%%%%%%%%%%%%%%%%%%%%%%%%%%%%%%%%%%%%%%%%%%%%%%%%%%%%%%%%
%\pagebreak
%\section*{Density Functional Theory}
%\section*{Electron microscopy and density functional theory}

%%%%%%%%%%%%%%%%%%%%%%%%%%%%%%%%%%%%%%%%%%%%%%%%%%%%%%%% FIGURE 3 %%%%%%%%%%%%%%%%%%%%%%%%%%%%%%%%%%%%%%%%%%%%%%%%%%%%%%%%%
\begin{figure}
\includegraphics[width=0.9\textwidth]{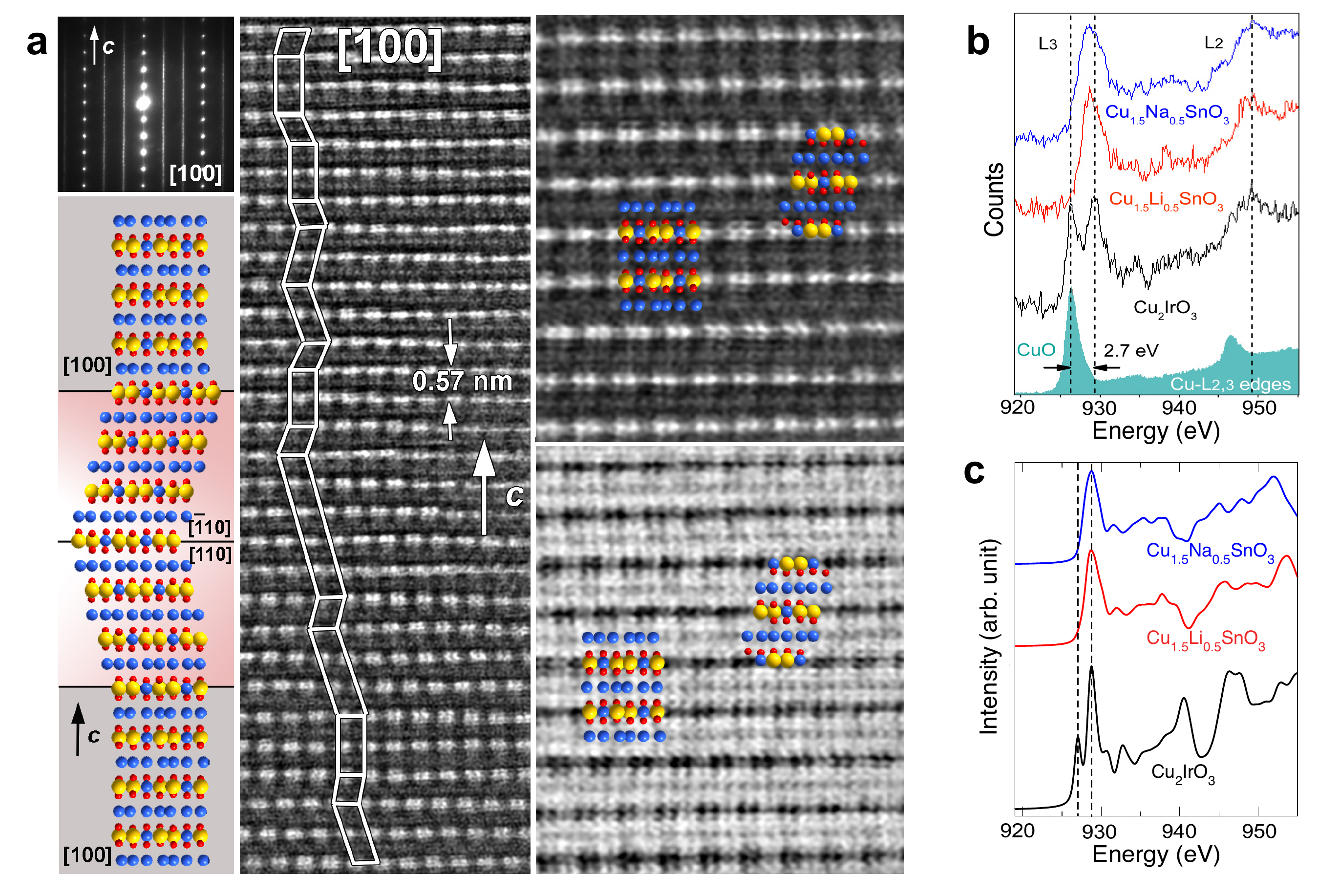}
\caption{\label{TEM}
\textbf{$\vert$~TEM and EELS data.}
\small{
\textbf{a,} Scanning transmission electron microscopy (STEM) is used to reveal perfect honeycomb ordering and twinned stacking disorder in \CIO.
Top left image is an electron diffraction pattern along $[100]$ where the streaking reveals stacking disorder along $c$-axis.
Middle panel shows HAADF-STEM image along $[100]$ with a zigzag stacking that is modeled in the left inset as a twinning between $[100]$, $[110]$, and $[\bar{1}10]$ directions.
Yellow, blue, and red circles represent Ir, Cu, and O atoms, respectively.
Right top and bottom panels are magnified HAADF-STEM and ABF-STEM images, respectively.
%
%In each image, one unit cell along $[100]$ and one unit cell along $[\bar{1}10]$ are modeled.
%
%Perfect honeycomb ordering is observed within each layer.
%
\textbf{b,} Experimental EELS spectra are compared between the stannates, \CLSO\ and \CNSO, and the iridate \CIO.
Only one $L_3$ peak is observed in the stannates corresponding to Cu$^+$ (note the CuO reference).
\CIO\ shows two $L_3$ peaks corresponding to Cu$^+$ and Cu$^{2+}$.
\textbf{c,} Self-consistent DFT calculations reproduce EELS spectra in agreement with the experiments.
The calculations reveal one peak in stannates corresponding to Cu$^+$ in dumbbell coordination but two peaks in \CIO\ due to mixed valence of copper (see Supplementary information for details of DFT calculations).
}
}
\end{figure}
%%%%%%%%%%%%%%%%%%%%%%%%%%%%%%%%%%%%%%%%%%%%%%%%%%%%%%%%%%%%%%%%%%%%%%%%%%%%%%%%%%%%%%%%%%%%%%%%%%%%%%%%%%%%%%%%%%%%%%%%%%%

%%
The most fundamental ingredient of a Kitaev material, apart from having spin-$1/2$ ions, is the honeycomb geometry.
A direct image of \CIO\ lattice is presented in Fig.~\ref{TEM}a obtained by transmission electron microscopy (TEM).
The middle panel is a high angle annular dark field scanning TEM image (HAADF-STEM) viewing down the $[100]$ axis of a small crystallite.
It reveals a zigzag stacking pattern along the $c$-axis that is modeled in the left inset as a rotation (twinning) between adjacent layers with alternating $[100]$, $[110]$, and $[\bar{1}10]$ orientations.
A similar twinned stacking disorder, i.e. $\pm 60^\circ$ rotation between adjacent layers, is observed in related stannate materials, \CLSO\ and \CNSO\ with alkali/tin honeycomb layers~\cite{abramchuk_crystal_2018}.
In the right upper and lower insets of Fig.~\ref{TEM}a, unit cell models with $[100]$ and $[\bar{1}10]$ orientations are overlaid on magnified views of the HAADF-STEM and ABF-STEM (annular bright field scanning TEM) images, respectively.
In both images, the layers exhibit a flawless pattern of Ir pairs separated by individual Cu atoms which is characteristic of honeycomb ordering~\cite{abramchuk_crystal_2018}.
Therefore, despite a twinned stacking disorder, each individual layer in \CIO\ has perfect honeycomb ordering without site mixing or vacancies.% -- a necessary condition for the Kitaev model.
TEM is also used for electron energy loss spectroscopy (EELS) with the data presented in Fig.~\ref{TEM}b.
A comparison between the $L_3$-edge in stannates and \CIO\ confirms that \CIO\ contains both Cu$^+$ and Cu$^{2+}$ whereas the stannates contain only Cu$^+$.
In the stannate materials, Cu atoms are restricted between the honeycomb layers in a dumbbell coordination~\cite{abramchuk_crystal_2018}.
Thus, all Cu$^{2+}$ in \CIO\ must be contained within the layers.
Self-consistent DFT calculations in Fig.~\ref{TEM}c reproduce the EELS spectra and confirm a single $L_3$ peak in stannates but two distinct peaks in \CIO.
%%

%%%%%%%%%%%%%%%%%%%%%%%%%%%%%%%%%%%%%%%%%%%%%%%%%%%%%%%%%%%%%%%%%%%%%%%%%%%%%%%%%%%%%%%%%%%%%%%%%%%%%%%%%%%%%%%%%%%%%%%%%%%
%%%%%%%%%%%%%%%%%%%%%%%%%%%%%%%%%%%%%%%%%%%%%%%%%%%%%%%%%% SUMMARY %%%%%%%%%%%%%%%%%%%%%%%%%%%%%%%%%%%%%%%%%%%%%%%%%%%%%%%%
%%%%%%%%%%%%%%%%%%%%%%%%%%%%%%%%%%%%%%%%%%%%%%%%%%%%%%%%%%%%%%%%%%%%%%%%%%%%%%%%%%%%%%%%%%%%%%%%%%%%%%%%%%%%%%%%%%%%%%%%%%%
%\pagebreak
%\section*{Summary}
%%
The emerging picture from our experimental and theoretical results is as follows.
\CIO\ contains a majority/minority of Cu$^+$/Cu$^{2+}$.
The minority Cu$^{2+}$ comprises about $1/3$ to $1/2$ of the copper ions within the honeycomb layers and nucleate regions of static magnetism with short-range correlations.
The majority of Cu$^+$ ions render a majority of Ir$^{4+}$ ions with $J_{\textrm {eff}}=1/2$ within the layers forming regions of Kitaev spin liquid phase.
Muons could implant either in the static magnetic domains and exhibit fast depolarization or in the spin liquid domains and exhibit slow depolarization.
The competition between static and dynamic magnetism revealed by \msr\ highlights the robustness of the spin liquid phase in \CIO\ and its ability to compete with static magnetism on equal footing.
%The collection of our \msr, XANES, DFT and TEM data unveil a competition between static and dynamic magnetism in the ground state of copper iridate.
%%

%%%%%%%%%%%%%%%%%%%%%%%%%%%%%%%%%%%%%%%%%%%%%%%%%%%%%%%%%%%%%%%%%%%%%%%%%%%%%%%%%%%%%%%%%%%%%%%%%%%%%%%%%%%%%%%%%%%%%%%%%%%
%%%%%%%%%%%%%%%%%%%%%%%%%%%%%%%%%%%%%%%%%%%%%%%%%%%%%%%%%% METHODS %%%%%%%%%%%%%%%%%%%%%%%%%%%%%%%%%%%%%%%%%%%%%%%%%%%%%%%%
%%%%%%%%%%%%%%%%%%%%%%%%%%%%%%%%%%%%%%%%%%%%%%%%%%%%%%%%%%%%%%%%%%%%%%%%%%%%%%%%%%%%%%%%%%%%%%%%%%%%%%%%%%%%%%%%%%%%%%%%%%%
\pagebreak
\begin{methods}
\subsection{Material Synthesis.}
\CIO\ was synthesized using a topotactic cation exchange reaction according to \NIO\ + $2$CuCl $\to$ \CIO\ + $2$NaCl under mild conditions (350~\C\ and 16~h).
Details of the synthesis are explained in reference~\cite{abramchuk_cu2iro3:_2017}.
\subsection{Muon Spin Relaxation.}
\msr\ measurements were performed at the ISIS Pulsed Neutron and Muon Source at the Rutherford Appleton Laboratories (UK) using the EMU and MuSR spectrometers with the sample inside a dilution refrigerator and a helium exchange cryostat, respectively.
The powder sample was pressed into a disk of 8~mm diameter and 1.9~mm thickness, and was wrapped in a 12.5 $\mu m$ thin silver foil.
Measurements in EMU were performed on a silver mounting pedestal in a Dilution fridge (50~mK$<T<$4.5~K, along with data at 16.4~K).
Due to the small sample area, measurements inside the dilution refrigerator were made in flypast mode (SI reference) in order to reduce the signal from muons not landing in the sample.
%~\cite{lynch_measuring_2003}
In this case, the background results from muons landing in the cryostat.
Measurements in the MuSR spectrometer were performed with the same sample mounted on a silver mounting plate in a helium exchange cryostat (1.7~K$<T<$~20~K).
In this case the background results from muons landing in the silver holder.
The background signals for each spectrometer were fixed at the values determined from long-time asymmetry at low temperatures (40\% of the total signal for EMU, 76\% for MuSR), where the sample was strongly magnetic.
The total asymmetry was fixed at the value determined from the initial asymmetry at high temperatures where the material had no fast relaxing component.
The sample contribution to the asymmetry is the difference between these two values.
%~\cite{pratt_wimda:_2000}
Data were fit using WIMDA software (SI reference) and all fits had a $\chi^2$ per degree of freedom of approximately $1.01$.
The fitting parameter $\alpha$, which quantifies the efficiency mismatch between front and back detectors~\cite{Yaouanc_muon_2011}, was determined by the application of a weak transverse magnetic field.
\subsection{X-ray absorption near edge spectroscopy (XANES)}
XANES measurements were performed at the Materials Research Collaborative Access Team (MRCAT), Sector 10-BM beam line at Argonne National Laboratory's Advanced Photon Source. %~\cite{kropf_new_2010}. 
Between 2 and 5~mg of  A$_2$IrO$_3$ (A~=~Li,~Na,~Cu) as well as IrO$_2$ powders were thoroughly ground with BN as a filler and PVDF (polyvinylidene fluoride) as a binder, pressed into a 5~mm diameter pellet, and encapsulated in thin Kapton tape.
Low temperature measurements were taken in transmission mode using a liquid nitrogen cooled stage (Linkam Scientific) at the Ir L$_3$-edge and the Cu $K$-edge.
The XANES data were reduced using the Athena program and fitted to structural models using the Artemis program, both of the IFEFFIT suite (SI reference). %~\cite{newville_ifeffit_2001,ravel_athena_2005}.
Ir data were fitted with a single Ir--O path using a range of 2--12 \AA$^{-1}$ ($dk=4$~\AA$^{-1}$) in $k$-space and 1--2~\AA\ ($dR=0.2$~\AA) in $R$-space and a weighting factor of $k^2$.
Cu data were fitted using the same ranges but with multiple weighting factors of $k$, $k^2$, and $k^3$~\cite{rehr_ab_2009}.
Cu data were fitted with a single Cu--O path as well as multiple Cu--O paths.
Cu $L$-edge data were collected at room temperature in total electron yield mode at the IEX beamline, 29-ID of the Advanced Photon Source, Argonne National Laboratory.
The beamline resolution was 250 meV.
\subsection{Electron Microscopy.}
Transmission electron microscopy (TEM) including electron diffraction (ED) and high angle annular dark field scanning TEM (HAADF--STEM), annular bright field scanning TEM (ABF--STEM), and electron energy loss spectroscopy (EELS) experiments were performed using an aberration double--corrected JEM ARM200F microscope operated at 200~kV and equipped with a CENTURIO EDX detector, Orius Gatan CCD camera and GIF Quantum spectrometer.
TEM samples were prepared by grinding the materials in an agate mortar with ethanol and depositing the obtained suspension on a Ni--carbon holey grid.
\subsection{Density Functional Theory.}
%%~\cite{kresse_efficiency_1996}~\cite{perdew_generalized_1996}~\cite{dudarev_electron-energy-loss_1998}
The geometric optimization of \CIO, \CNSO, and \CLSO\ were implemented in the pseudopotential VASP code (SI reference) using a projected augmented wave (PAW) method and the Perdew-Burke-Ernzerhof (PBE) exchange-correlation potential (SI reference).
The Hubbard correction was implemented using the Dudarev's scheme (SI reference) with $U_{\textrm{eff}}=3$~eV for iridium $5d$ orbitals and $5$~eV for copper $3d$ orbitals.
The atomic positions were relaxed until forces were converged to $0.03$~eV/\AA.
%
%Crystal structures were visualized using the VESTA program~\cite{momma_vesta_2011}.
%~\cite{blaha_wien2k_2018}
Simulations of the spectroscopic data were implemented in the full potential Wien2k code (SI reference) using a linearized augmented plane wave (LAPW) approach and PBE0 hybrid functional with on-site corrections to iridium $5d$ and copper $3d$ orbitals.
Radius of muffin tin (RMT) was selected to be $1.46, 1.48, 1.50, 2.00, 2.00, 1.94$ bohr for O, Li, Na, Ir, Sn, and Cu atoms and the basis size control parameter was $RK_{\textrm{max}}=6$.
Both structural relaxation and spectroscopic calculations were spin polarized and included spin orbit coupling (SOC).

\end{methods}

%%%%%%%%%%%%%%%%%%%%%%%%%%%%%%%%%%%%%%%%%%%%%%%%%%%%%%%%%%%%%%%%%%%%%%%%%%%%%%%%%%%%%%%%%%%%%%%%%%%%%%%%%%%%%%%%%%%%%%%%%%%
%%%%%%%%%%%%%%%%%%%%%%%%%%%%%%%%%%%%%%%%%%%%%%%%%%%%%%%%% REFERENCE %%%%%%%%%%%%%%%%%%%%%%%%%%%%%%%%%%%%%%%%%%%%%%%%%%%%%%%
%%%%%%%%%%%%%%%%%%%%%%%%%%%%%%%%%%%%%%%%%%%%%%%%%%%%%%%%%%%%%%%%%%%%%%%%%%%%%%%%%%%%%%%%%%%%%%%%%%%%%%%%%%%%%%%%%%%%%%%%%%%

% Put the bibliography here, most people will use BiBTeX in
% which case the environment below should be replaced with
% the \bibliography{} command.

\pagebreak
\bibliographystyle{nature} % choose style first
\bibliography{Kenny_1nov2018}
%\nocite{*}

%%%%%%%%%%%%%%%%%%%%%%%%%%%%%%%%%%%%%%%%%%%%%%%%%%%%%%%%%%%%%%%%%%%%%%%%%%%%%%%%%%%%%%%%%%%%%%%%%%%%%%%%%%%%%%%%%%%%%%%%%%%
%%%%%%%%%%%%%%%%%%%%%%%%%%%%%%%%%%%%%%%%%%%%%%%%%%%%%%%%% ADDENDUM %%%%%%%%%%%%%%%%%%%%%%%%%%%%%%%%%%%%%%%%%%%%%%%%%%%%%%%
%%%%%%%%%%%%%%%%%%%%%%%%%%%%%%%%%%%%%%%%%%%%%%%%%%%%%%%%%%%%%%%%%%%%%%%%%%%%%%%%%%%%%%%%%%%%%%%%%%%%%%%%%%%%%%%%%%%%%%%%%%%

%% Here is the endmatter stuff: Supplementary Info, etc.
%% Use \item's to separate, default label is "Acknowledgements"
\pagebreak
\begin{addendum}
 \item[Acknowledgments] We are grateful to Y.~Ran for fruitful discussions.
 F.T. and M.A. acknowledge support from the National Science Foundation, Award No. DMR--1708929.
 MRCAT operations are supported by the Department of Energy and the MRCAT member institutions.
 This research used resources of the Advanced Photon Source, a U.S. Department of Energy (DOE) Office of Science User Facility operated for the DOE Office of Science by Argonne National Laboratory under Contract No. DE-AC02-06CH11357.
 Experiments at the ISIS Pulsed Neutron and Muon Source were supported by a beamtime allocation from the Science and Technology Facilities Council.
 O.I.L acknowledges financial support from the "Agence Nationale de la Recherche" in the framework of the "Investissements d'avenir" program with the reference "ANR--11--EQPX--0020" for EELS data obtained using GIF Quantum.
 W.L.-D.-H. and X.R. thank the HPC resources from GENCI-[TGCC/CINES/IDRIS] (Grant 2017-A0010907682).
 Work at APS was supported by the US Department of Energy (DOE), Office of Science, under Contract No. DE-AC02-06CH11357.
 \item[Author Contributions] F.T. and M.J.G. designed the experiment and wrote the paper.
 E.M.K, A.B., S.P.C., and M.J.G. performed \msr\ experiments and analyzed data.
 C.U.S., J.L.M., G.F., and D.H. performed XANES experiments and analyzed data.
 W.L.-D.-H. and X.R. performed DFT calculations.
 G.S. analyzed data.
 M.A. and F.B. synthesized the material and performed X-ray refinements.
 O.I.L. and N.E.M. performed TEM and EELS experiments.
 All authors revised the manuscript.
 \item[Competing interests] The authors declare no competing interests.
 \item[Supplementary information] is available online.
 \item[Correspondence and requests for materials] should be addressed to F.T.~(email: fazel.tafti@bc.edu).
\end{addendum}

\end{document}